\documentclass[10pt,twocolumn]{article} 

\usepackage{simpleConference}
\usepackage{times}
\usepackage{graphicx}
\usepackage{amssymb}
\usepackage{url,hyperref}
\usepackage[square,sort,comma,numbers]{natbib}

\usepackage{lineno,hyperref}
\usepackage[utf8]{inputenc}
\usepackage{float}
\usepackage[table]{xcolor}
\usepackage{enumitem}
\usepackage{pgfplots}
\usepackage{amssymb}
\usepackage{pifont}
\newcommand{\cmark}{\ding{51}}%
\newcommand{\xmark}{\ding{55}}%
\pgfplotsset{width=7cm} 

\usepackage{tabularx}
\usepackage{rotating}
\usepackage{diagbox}

\begin{document}

\title{Software Testing Process Models Benefits \& Drawbacks:\\ a Systematic Literature Review}

\author{Katarína Hrabovská, 
Bruno Rossi, and 
Tomáš Pitner
\\Faculty of Informatics\\Masaryk University, Brno, Czech Republic
}



\twocolumn[
  \begin{@twocolumnfalse}
     \maketitle

\begin{abstract}
\textit{Context:} Software testing plays an essential role in product quality improvement. For this reason, several software testing models have been developed to support organizations. However, adoption of testing process models inside organizations is still sporadic, with a need for more evidence about reported experiences.

\noindent \textit{Aim:} Our goal is to identify results gathered from the application of software testing models in organizational contexts. We focus on characteristics such as the context of use, practices applied in different testing process phases, and reported benefits \& drawbacks.

\noindent \textit{Method:} We performed a Systematic Literature Review (SLR) focused on studies about the application of software testing processes, complemented by results from previous reviews.

\noindent \textit{Results:} From 35 primary studies and survey-based articles, we collected 17 testing models. Although most of the existing models are described as applicable to general contexts, the evidence obtained from the studies shows that some models are not suitable for all enterprise sizes, and inadequate for specific domains.

\noindent \textit{Conclusion:} The SLR evidence can serve to compare different software testing models for applicability inside organizations. Both benefits and drawbacks, as reported in the surveyed cases, allow getting a better view of the strengths and weaknesses of each model.
\end{abstract}

  \end{@twocolumnfalse}
]

\section{Introduction}
Software testing is a key phase of the software development process, as it represents the process of quality validation and verification of a software product \cite{ref:young2008software-testing}. Such phase is even more crucial nowadays, as software has become increasingly complex, mission-, safety- critical, and essential in daily activities, calling for an increase in quality~\cite{ref:dustin2002testing,ref:movzucha2016mutation,ref:rossi2010modelling}.

Testing models and standards emerged over time to more clearly define a testing process with formalized criteria to make the whole process repeatable and reproducible~\cite{ref:matalonga2015matching}. Testing standards play a supportive role in reaching the desired quality level, allowing for a more systematic and structured way for further analysis and improvements~\cite{burnstein_developing_1996}.

Over time, several software testing standards and models have emerged to fulfill various requirements of organizations for a testing process \citep{garcia_test_2014,afzal_software_2016,garousi_software_2017}. The number of models can allow organizations to adopt the one that mostly satisfies organizational needs, even though full adoption of testing standards by organizations has been found to be sporadic~\citep{ref:kasurinen2011test,ref:garousi2013survey}. 

The goal of this paper is to identify existing testing models and their characteristics such as domain, context of application, aspects improved, advantages and drawbacks, important models' practices in the industrial context, by complementing existing reviews of Garcia et al.~\cite{garcia_test_2014}, Afzal et al.~\cite{afzal_software_2016}, and Garousi et al.~\cite{garousi_software_2017}. Information from the review can help practitioners in selecting the most appropriate model for a context. 

There is a huge variety of methods that can be applied for literature reviews in the context of Evidence Based Software Engineering (EBSE) \citep{ref:kitchenham2004evidence,ref:dyba2005evidence}. In this paper, we conduct a Systematic Literature Review (SLR)  \citep{ref:kitchenham2004procedures,ref:kitchenham2009systematic}. The review contributes to the area of testing models by identifying and analyzing the most popular models published in empirical studies. Compared to previous reviews in the area (\cite{garcia_test_2014,afzal_software_2016,garousi_software_2017}), our focus is more on experiences from empirical studies, by providing models characteristics which can be crucial for organizations in the selection process of an appropriate model. In our review, we focus on the following contributions:

\begin{itemize}
  \item the list of the most used testing models applied in empirical studies published in the literature, with the indication of the application domains;
  \item the main aspects supported by the testing models and on which organizations are focused when adopting the model;
  \item the identification of strengths and weaknesses of the testing models and their practices inside organizations as reported by the included cases;
\end{itemize}

The paper is structured as follows. Section 2 explains key terminology of testing standards and results from other existing surveys. Section 3 describes the entire research method and specific steps of the process. Section 4 presents the results of the review answering the research questions. Section 5 discusses the contribution of the review and compare the results with other existing surveys. Section 6 presents possible threats to validity of the review. Section 7 presents the conclusions.
 
\section{Background}
Software testing is an important phase of software development. IEEE definition \citep{ref:IEEE610} suggests that \textit{"software testing is the process of exercising or evaluating a system by manual or automatic means to verify that it satisfies specified requirements or to identify differences between actual and expected results"}. Software testing is crucial as it involves critical activities from unit, integration, system, to acceptance testing \citep{ref:young2008software-testing}.

Following indications from testing models / standards and best practices, it is essential to improve the overall testing process. Nevertheless, there are still many organizations that do not follow best practices and perform testing in an ad-hoc way \citep{ref:kasurinen2011test, bertolino_software_2007}. 

For the purpose of facilitating and improving testing in organizations, several test models have been designed since the '80s \citep{garousi_software_2017}. The aim of the models is to develop structured and systematic testing processes based on organizational needs. 

In literature, there are different terms used when addressing formalization of testing processes, such as \textit{testing frameworks}, \textit{testing improvement processes}, \textit{testing models}, \textit{testing standards}, with the following definitions: 

\begin{itemize}
\item The term \textit{testing framework} is more often used in the sense of specific tools and practices applied for tests creation~\citep{ref:eisenbiegler_2013}.
\item \textit{Software testing models} and \textit{test improvement processes} are commonly used interchangeably. However, according to Veenendaal~\cite{ref:veenendaal_2016}, two types of models are distinguished for test process improvement: \textit{test improvement models} (also called \textit{test maturity models}) and \textit{content-based models}.
\begin{itemize}
\item The core of \textit{test improvement models} consists of testing best practices with the structure divided into several maturity levels~\citep{ref:veenendaal_2016}.
\item \textit{Content-based models} describe different testing practices in detail without pre-defined paths for the improvement process in the form of different maturity levels~\citep{ref:veenendaal_2016}.
\end{itemize}
\end{itemize}

According to the ISO/IEC/IEEE 24765:2017(E) standard~\cite{ISO_terminology}, models and standards are defined as follows:

\begin{itemize}
\item a model is the \textit{``representation of a real-world process, device, or concept''}~\cite{ISO_terminology};
\item a standard is a \textit{"document, established by consensus and approved by a recognized body, that provides, for common and repeated use, rules, guidelines or characteristics for activities or their results, aimed at the achievement of the optimum degree of order in a given context"}~\cite{ISO_terminology}.
\end{itemize}

In this paper, we focus on the whole set of testing models, including both categories: test maturity models and content-based models. We also include testing standards (e.g. ISO/IEC/IEEE 29119 \citep{ali_formalizing_2015,kasurinen_self-assessment_2011}), as they are testing models backed by standards development organizations like IEEE, IEC, and ISO. From this point, when we refer to testing models, we include test improvement models, content-based models and testing standards.

\subsection{Testing Models}
The development of testing models started based on the lack of focus on testing aspects by software process improvements (SPI) models. SPI models support organizations in quality improvement of their services and products by optimizing processes of software development. To achieve this aim, several models and standards such as the Capability Maturity Model (CMM), Capability Maturity Model Integrated (CMMI) and the Software Process Improvement and Capability Determination (SPICE) have been developed over time \citep{galinac_empirical_2009}.

One of the distinguishing features of models is their architecture. One group forms processes with staged architectures such as CMMI, where improvements are applied based on capability and maturity levels. On the other hand, standards such as ISO 9001 suggest the architecture, which specifies requirements of a quality management system \citep{afzal_software_2016}.   

SPI models focus on several aspects of quality improvement. However, software testing constitutes only a small part of SPI and for that reason, a number of testing improvement models has been defined specifically for testing purposes. Many of them are supplementary to SPI processes because of the similarities in their structures. Testing models help organizations to suggest and evaluate improvements of testing processes. The most popular testing approaches are Testing Maturity Model (TMM), Test Maturity Model integration (TMMi) and Test Process Improvement (TPI); moreover, they form the basis of several other existing testing standards \citep{garcia_test_2014}.
 
\begin{table*}[!htb]
\centering
\renewcommand{\arraystretch}{1.8}
\caption{Contributions of existing testing models reviews}
\label{table:1}
\begin{footnotesize}
\begin{tabular}{ p{3.0cm} p{1cm} p{5cm} p{1.4cm} p{5cm} } 
 \hline
 \bf{Reference} & \bf{Year} & \bf{Title} & \bf{Reviewed models} & \bf{Notes} \\ 
 \hline
 Garcia et al.~\cite{garcia_test_2014} & 2014 & Test Process Models: Systematic Literature Review & 23 & \textbf{Focus:} Identify testing process models defined over the years. \textbf{Method:} SLR  \\
 Afzal et al.~\cite{afzal_software_2016} & 2016 & Software test process improvement approaches: A systematic literature review and an industrial case study & 18 & \textbf{Focus:} usability of the models in industry. \textbf{Method:} SLR + case study (TPI NEXT and TMMi) \\
 Garousi et al.~\cite{garousi_software_2017} & 2017 & Software test maturity assessment and test process improvement: A multivocal literature review & 58 & \textbf{Focus:} consolidate the list of all test maturity models proposed by practitioners and researchers. \textbf{Method:} MLR (included gray literature)\\
 \hline
\end{tabular}
\end{footnotesize}
\end{table*}

\begin{table*}[!htb]
\centering
\renewcommand{\arraystretch}{1.8}
\caption{Contributions of existing testing models reviews (reviews focused on selected groups of models)}
\label{table:2}
\begin{footnotesize}
\begin{tabular}{ p{3.0cm} p{1cm} p{5cm} p{1.4cm} p{5cm} } 
 \hline
 \bf{Reference} & \bf{Year} & \bf{Title} & \bf{Reviewed models} & \bf{Notes} \\ 
 \hline
 Swinkels~\cite{swinkel_software_2000} & 2000 & A Comparison of TMM and other Test Process Improvement Models & 8 & \textbf{Focus:} provide description of all available TPI models up to 2000; compare the TMM model with selected TPI models; and obtain input for the development of a new model MB-TMM. \textbf{Method:} Informal survey + comparison of selected models (as a part of Technical Report) \\
 Farooq and Dumke~\cite{ref:farooq2008review} & 2008 & Evaluation Approaches in Software Testing & 6 & \textbf{Focus:} evaluation of testing models and brief comparison of selected testing models (TMM, TPI and TMMi). \textbf{Method:} Informal comparison (as a part of Technical Report) \\
 Farooq~\cite{ref:farooq2009review} & 2009 & An Evaluation Framework for Software Test Processes & 8 & \textbf{Focus:} evaluation of testing models and comparison of selected testing models (TMM, TPI, ICMM and TMMi). \textbf{Method:} Informal comparison (as a part of Dissertation) \\
Abdou et al.~\cite{ref:abdou2013review} & 2013 & Managing Corrective Actions to Closure in Open Source Software Test Process & 4 & \textbf{Focus:} provide overview of main features of four maturity models (TMM, TIM, TPI and TMMi) and propose a framework for software testing assessment based on TMMi. \textbf{Method:} Informal comparison \\
 \hline
\end{tabular}
\end{footnotesize}
\end{table*}

\subsection{Results from other reviews}
Several secondary studies of testing process models were performed with the aim to compare existing models \cite{garousi_software_2017}.

We compare results and used methods of three literature reviews about testing models conducted in recent years to analyze existing approaches, namely by Garcia et al.~\cite{garcia_test_2014}; Afzal et al.~\cite{afzal_software_2016}; Garousi et al.~\cite{garousi_software_2017} (Table \ref{table:1}). The current review complements existing reviews by focusing explicitly on empirical studies, extracting the list of applied models and their domains, aspects, strengths and weaknesses as derived from empirical studies. Provision of this knowledge can help organizations to analyze and decide which testing models can be more appropriate for specific contexts.

The review of Garcia et al.~\cite{garcia_test_2014} presents 23 testing models and information about them such as domain, source model and year of publication. It includes models such as the Maturity Model for Automated Software Testing (MMAST), Testing Assessment Programme (TAP) and Test Organization Model (TOM), which are assessed rarely in empirical studies. These models are collected from the report of Swinkels~\cite{swinkel_software_2000}, which compares the models against TMM.  

\noindent \textbf{Differences with the current review.} The main difference between the survey of Garcia et al.~\cite{garcia_test_2014} and our review is about the focus of the review. Garcia et al.~\cite{garcia_test_2014} aim at getting a complete list of models created over time, while we concentrate on experiences derived from the application of the models to real organizational context. As such, in our review, advantages, drawbacks and practices reported for the models play a major role. Our review is also more extensive in terms of provided details about testing models than the one of Garcia et al.~\cite{garcia_test_2014}.

The second review by Afzal et al.~\cite{afzal_software_2016} provides both an SLR and a case study by mapping TPI NEXT and TMMi to a concrete case. The specific focus is on usability of the models in industry. Overall the study provides a list of 18 testing models. There are less approaches than in the previous survey by Garcia et al.~\cite{garcia_test_2014}, because testing models from secondary studies were omitted, e.g. Swinkels~\cite{swinkel_software_2000}. On the other hand, the review by Afzal et al.~\cite{afzal_software_2016} is more extended in models description and details such as the status of development, and improvement suggestions.

\noindent \textbf{Differences with the current review.} The differences between the review of Afzal et al.~\cite{afzal_software_2016} and our survey are mainly in the approach. Afzal et al.~\cite{afzal_software_2016} look into the applicability of the testing models to industry, by reviewing list of models in an SLR. However, their focus is not on the identification of reported benefits and drawbacks, rather on the identification of models applicable to industry. Such models are then used in a case study to assist organizations in the selections of best testing models. In our case, we look at reported experiences directly from the articles included in the review.

The most recent survey is a Multivocal Literature Review (MLR) by Garousi et al.~\cite{garousi_software_2017}, which focuses on scientific literature and also on gray literature. As a result, the authors identified 58 testing models. They collected the drivers, benefits and challenges of improvement activities. The survey deals with base maturity models which are used for developing new models. 

\noindent \textbf{Differences with the current review.} Besides the used methods, the review of Garousi et al.~\cite{garousi_software_2017} differs from our survey in the focus of the study. Garousi et al.~\cite{garousi_software_2017} place more relevance to challenges and benefits of conducting testing process improvement activities inside organizations, as such the review of Garousi et al.~\cite{garousi_software_2017} is more detailed than the current one. While the authors discuss drivers and challenges for all models in general, we emphasize more the empirical experiences gathered from the application of each single testing model, associating reported benefits, drawbacks and practices to drive adoption decisions about each testing model. 

\medskip
There are other secondary studies that focus on selected groups of testing models, although such studies are not run in the form of an SLR (Table \ref{table:2}). 

The technical report of Swinkels~\cite{swinkel_software_2000} gathers existing TPI models up to year 2000. The result is eight testing models (TMM, MMAST, TAP, TCMM, TIM, TOM, TPI and TSM). Subsequently, the author defines criteria for models selection and compares the TMM model only with those which fulfill the criteria (TIM and TPI). As a conclusion from the comparison, authors consider TMM and TPI as the most comprehensive models and propose TMM as a base for MB-TMM, with several key areas handled by TPI.

The report by Farooq and Dunke~\cite{ref:farooq2008review} includes a part of comparison of several testing models. The authors present six testing models (TMM, TPI, TIM, MB-VV-MM, TPAM and TMMi) and select three of them, which are discussed in more extended way (TMM, TPI and TMMi). 

Farooq~\cite{ref:farooq2009review} evaluates eight testing models in his dissertation (TMM, TPI, TIM, TMap, MB-VV-MM, TPAM, ICMM and TMMi). Because of the insignificance or incompleteness of some models, the author selects four models (TMM, TPI, ICMM and TMMi) and provides their overview. 

The paper by Abdou et al.~\cite{ref:abdou2013review} focuses on four testing maturity models (TMM, TIM, TPI and TMMi) with the aim of selecting a model for development of a framework for software testing assessment. The authors provide an overview of the main models' features. The second part of the paper consists of a proposal for the software testing assessment framework based on the TMMi model. 

\section{Systematic Literature Review}
\begin{figure*}[!htb]
\centering
\caption{SLR Search Strategy}
\label{figure:strategy}
	\includegraphics[width=0.7\linewidth]{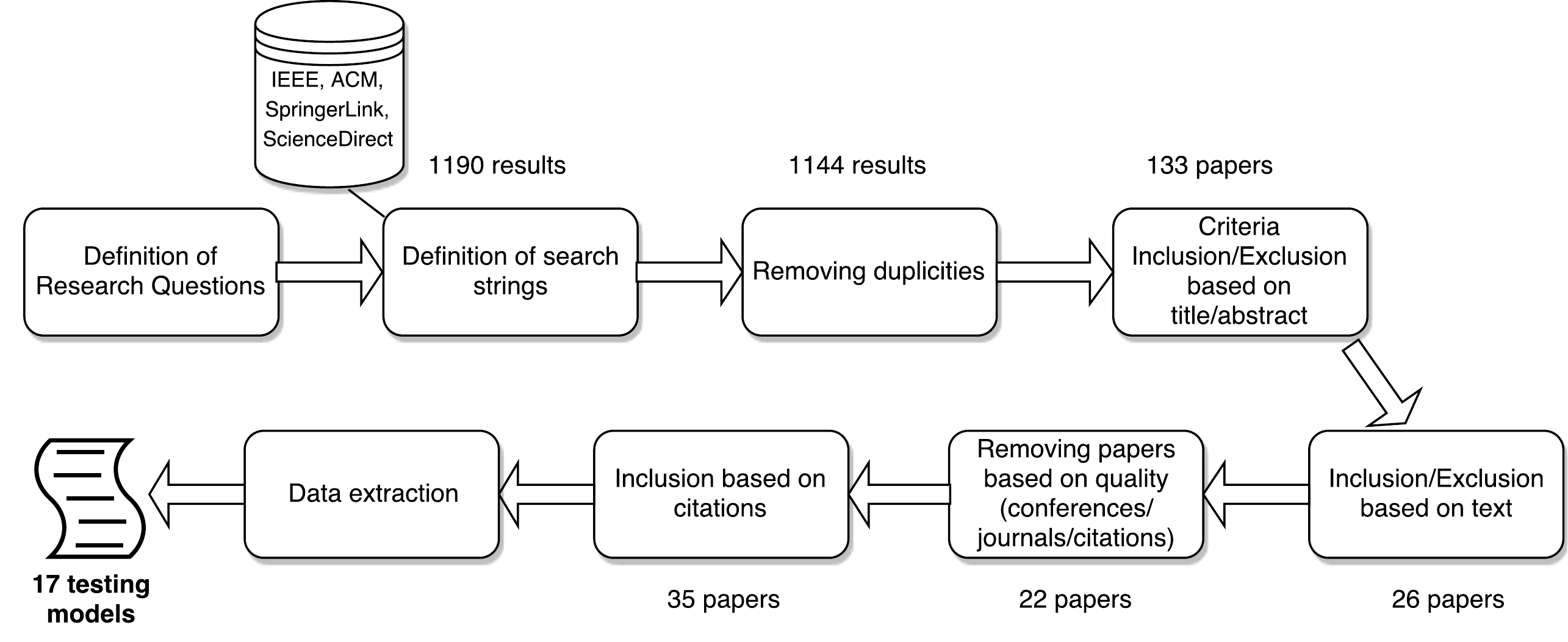}
\end{figure*}

To collect empirical evidence about existing testing models, we performed a Systematic Literature Review (SLR). An SLR is represented by a set of systematic steps, which gather and evaluate relevant literature with the aim to answer research questions. The whole SLR process is designed and performed in a more formalized way than traditional literature reviews \citep{ref:kitchenham2009systematic}. Following the guidelines for conducting SLRs provided by Kitchenham~\cite{ref:kitchenham2004procedures}, we defined the review protocol (Fig. \ref{figure:strategy}). First we codify several research questions on which we are focused during the review process. Afterwards we specify the search strategy including databases selection and queries preparation. We then define the inclusion and exclusion criteria and we run the queries on the selected digital repositories. Thereupon, we perform a quality evaluation step by establishing a further inclusion criteria based on citations ratio and rankings of conferences / journals where articles were published. To improve the completeness of the results, the next step is the inclusion of selected relevant papers that were included in other reviews and are relevant according to the goals of this review. Finally, we extract relevant data based on the research questions.

\subsection{Research Questions}
To reach the goal of identifying and analyzing existing testing standards, the following research questions were formulated:

\begin{itemize}
  \item [\bf{RQ1:}] \textit{Which are the main testing models and their characteristics reported in empirical studies?} Rationale: we collect models that were reported in empirical papers. This will allow to create an overview of the most used models and answer further research questions.
  \item [\bf{RQ2:}] \textit{Which are the reported aspects of a testing process the testing models help to improve?} Rationale: organizations can look into aspects that were reported to be improved after the application of the testing models.
  \item [\bf{RQ3:}] \textit{Which are the advantages and drawbacks of each testing model as reported in the empirical papers?} Rationale: identification of strengths and weaknesses of the testing models on the field can be useful for organizations to make a reasoned choice about the adoption.
  \item [\bf{RQ4:}] \textit{Which are the phases/practices of testing models for different domains?} Rationale: organizations can analyze which practices have the most value for them based on their domain.
\end{itemize}

\subsection{Search Strategy}
Keywords were defined based on the research questions to create search queries, including: \textit{maturity model, process improvement, improvement model, maturity standard, test standard, capability model, test process and test model}. Such keywords were adapted depending on the digital repositories (Table \ref{table:queries}). 

We used the search queries in four digital repositories: ACM Digital Library (https://dl.acm.org), IEEE Xplore Digital Library (https://ieeexplore.ieee.org), ScienceDirect (https://sciencedirect.com) and Springer Link (https://link.springer.com). These databases were selected because they include high quality scientific articles with generally low overlap in terms of duplications. On the other hand, Google Scholar was excluded because including already many of the articles from the other repositories. 

The search string was restricted to the title and some restrictions were applied for the specific repositories (Table \ref{table:queries}). In IEEE Xplore Digital Library the search string was extended by the \textit{'software'} keyword in the abstract, in Springer Link \textit{'software'} and \textit{'test'} were added to general searching and in ScienceDirect \textit{'software'} term was searched in the title, abstract and keywords. The table includes also the information about the number of results found in each database. The search process was performed in IEEE and ACM on 22nd October 2017 and in Springer Link and ScienceDirect on 24th October 2017. After merging the results and removing duplicities we obtained 1,144 papers.

\begin{table*}[!htb]
\centering
\renewcommand{\arraystretch}{1.8}
\caption{Search queries and number of results from the databases (queries run 22nd October 2017)}
\label{table:queries}
\begin{footnotesize}
\begin{tabular}{ p{3cm} p{10cm} p{1.5cm} } 
 \hline
 \bf{Database} & \bf{Search Query} & \bf{\# results} \\ 
 \hline 
 IEEEXplore & ("Document Title":"maturity model" OR "Document Title":"process improvement" OR "Document Title":"improvement model" OR "Document Title":"maturity standard" OR "Document Title":"test standard" OR "Document Title":"capability model" OR "Document Title":"test process" OR "Document Title":"test model" OR "Document Title":"testing standard" OR "Document Title":"testing process" OR "Document Title":"testing model") AND "Abstract":"software" & 479 \\
 ACM Digital Library & acmdlTitle:("maturity model" "process improvement" "improvement model" "maturity standard" "test standard" "capability model" "test process" "test model") & 253 \\
 Springer Link & software AND test
title contains: “maturity model” OR "process improvement" OR "improvement model" OR "maturity standard" OR "test* standard" OR "capability model" OR “test* process” OR “test* model” & 357 \\
ScienceDirect & TITLE({maturity model} OR {process improvement} OR {improvement model} OR {maturity standard} OR {test standard} OR {capability model} OR {test process} OR {test model} OR {testing standard} OR {testing process} OR {testing model}) and TITLE-ABSTR-KEY(software) & 101 \\
 \hline
\end{tabular}
\end{footnotesize}
\end{table*}

\subsection{Inclusion and exclusion criteria}
The set of resulting articles from the queries run on the digital repositories is relatively large (1,144). However, this set was reduced by the application of the inclusion and exclusion criteria. We defined the following inclusion criteria:

\begin{itemize}
	\item papers include any testing model or a comparison of different testing models;
    \item papers are empirical papers (case study, experiment, observation, survey-based papers);
\end{itemize}

As empirical papers we consider papers which are based on research of qualitative or quantitative approach and follow main research steps (Definition, Planning, Operation, Analysis \& interpretation, Conclusions and Presentation \& packaging) as described in the article of Wohlin et al.~\cite{ref:wohlin2012experimentation}. Similarly, we specified the exclusion criteria:

\begin{itemize}
  \item papers not in English;
  \item irrelevant publication types such as editorials, slides, posters and books;
\end{itemize}

The inclusion and exclusion criteria were applied during the filtering process, which was performed in several phases. The first phase included reading titles and abstracts, which significantly reduced the number of papers (133 papers remaining). The purpose of the next phase was to exclude articles based on reading the whole text. During this phase we removed 107 articles from the list and thus we finished the application of inclusion and exclusion criteria (26 paper remaining).

\subsection{Quality Evaluation}
The second part of the filtering process aimed at improving the quality of the results by setting some further criteria for inclusion of the selected papers. In this SLR, the main rationale was to consider number of citations of papers and the publication venue for articles, setting thresholds for the inclusion of the papers. However, we considered this as weak criteria: if a paper would fail to meet this criteria, it was still responsibility of the researchers to evaluate the exclusion of the paper.

To improve the quality of results, several articles were removed from the list based on a low number of citations per year (Table \ref{table:citations}). Two articles with low citation ratio were not eliminated from the list: Ali and Yue~\cite{ali_formalizing_2015} directly describe existing approach and references to the official 
ISO/IEC/IEEE 29119 Software Testing website (http://www.softwaretestingstandard.org/) and Lee~\cite{lee_adapting_2009} suggests a testing model and includes valuable information about the TMM model and its drawbacks. 

Subsequently, quality evaluation proceeded based on the conferences and journals where papers were published (Table \ref{table:conferences}). We collected information about ranking of the most conferences/journals, which are available online (http://www.conferenceranks.com/, http://portal.core.edu.au/conf-ranks/, http://portal.core.edu.au/jnl-ranks/). Based on the results, the article of Saiedian and Carr~\cite{saiedian_characterizing_1997} was removed because it was published in a newsletter.

\begin{table}[!htb]
\centering
\caption{Quality Evaluation based on citations ratio. Gray-marked papers were eliminated.}
\label{table:citations}
\begin{footnotesize}
\begin{tabularx}{\linewidth}{lrr} 
 \hline
 \bf{Article} & \bf{citations} & \bf{citations x year} \\ 
 \hline 
 Burnstein et al.~\cite{burnstein_developing_1996} & 201 & 9.1 \\
 Dyba~\cite{dyba_a_factors_2003} & 130 & 8.7 \\
 Paulish and Carleton~\cite{paulish_case_1994} & 127 & 5.3 \\
 Galinac~\cite{galinac_empirical_2009} & 23 & 2.6 \\
 Huo et al.~\cite{huo_exploratory_2006} & 25 & 2.1 \\
 Saiedian and Carr~\cite{saiedian_characterizing_1997} & 44 & 2.1 \\
 Wangenheim et al.~\cite{wangenheim_tailoring_2013} & 10 & 2.0 \\
 Ryu et al.~\cite{ryu_strategic_2008} & 19 & 1.9 \\
 Garcia et al.~\cite{garcia_adopting_2010} & 10 & 1.3 \\
 Jung~\cite{jung_test_2009} & 12 & 1.3 \\
 Furtado et al.~\cite{furtado_mpt.br:_2012} & 8 & 1.3 \\
 Oh et al.~\cite{oh_optimizing_2008} & 13 & 1.3 \\
 Camargo et al.~\cite{camargo_identifying_2013} & 7 & 1.2 \\
 Suwanya and Kurutach~\cite{suwanya_analysis_2008} & 11 & 1.1 \\
 Mirna et al.~\cite{mirna_advantages_2011} & 8 & 1.0 \\
 Rodrigues et al.~\cite{rodrigues_definiton_2010} & 8 & 1.0 \\
 Toroi et al.~\cite{toroi_identifying_2013} & 4 & 0.8 \\
 Nikula et al.~\cite{nikula_extending_2009} & 5 & 0.6 \\
 Simon~\cite{simon_spice:_1996} & 13 & 0.6 \\
 Rungi and Matulevičius~\cite{rungi_empirical_2013} & 2 & 0.4 \\
 Kumaresh and Baskaran~\cite{kumaresh_experimental_2012} & 3 & 0.4 \\
\rowcolor{gray!50!} Nieminen and Räty~\cite{nieminen_adaptable_2015} & 1 & 0.3 \\
 Lee~\cite{lee_adapting_2009} & 2 & 0.2 \\
\rowcolor{gray!50!} Senyard et al.~\cite{senyard_towards_2000} & 2 & 0.2 \\
 Ali and Yue~\cite{ali_formalizing_2015} & 0 & 0.0 \\
\rowcolor{gray!50!} Liu et al.~\cite{liu_research_2015} & 0 & 0.0 \\
 \hline
\end{tabularx}
\end{footnotesize}
\end{table}

\begin{table*}[!htb]
\centering
\caption{Quality Evaluation based on conferences/journals. Gray-marked line was eliminated from the list.}
\label{table:conferences}
\begin{footnotesize}
\begin{tabular}{ p{14cm} p{1.5cm} p{1.5cm} } 
 \hline
 \bf{Conference / Journal} & \bf{Rank} & \bf{\#} \\ 
 \hline 
 International Conference on Quality Software (QSIC) & B & 3 \\
 Health and Technology & A* & 1 \\
European Software Engineering Conference co-located ACM SIGSOFT International Symposium on Foundations of Software Engineering 	(ESEC/FSE) & A* & 1 \\
 IEEE International Conference on Software Maintenance (ICSM) & A & 1 \\
International Conference on Model Driven Engineering Languages and Systems (MODELS) & A & 1 \\
 Information and Software Technology & B & 1 \\
 Journal of Systems Architecture (JSA) & B & 1 \\
 International Test Conference (ITC) & B & 1 \\
 International Conference on the Quality of Information and Communications Technology (QUATIC) & C & 1 \\
International Workshop on Software Quality (WoSQ) & C & 1 \\
International Conference on Computer and Information Technology (ICCIT) & C & 1 \\
International Conference on Interaction Sciences: Information Technology, Culture and Human (ICIS) & C & 1 \\
International Conference on Computer and Information Science (ICIS) & C & 1 \\
 Australian Software Engineering Conference (ASWEC) & - & 1 \\
International Conference on Software Engineering Research, Management and Applications (SERA) & - & 1 \\
Electronics, Robotics and Automotive Mechanics Conference (CERMA) & - & 1 \\
International Conference on Recent Advances in Computing and Software Systems (RACSS) & - & 1 \\
Brazilian Symposium on Software Engineering (SBES) & - & 1 \\
International Conference on Information and Software Technologies (ICIST) & - & 1 \\
IEEE Computer (magazine) & - & 1 \\
 \rowcolor{gray!50!} Newsletter ACM SIGICE Bulletin & - & 1 \\
 \hline
\end{tabular}
\end{footnotesize}
\end{table*}

\subsection{Inclusion based on citations and references}
As the next step of the search strategy, we analyze papers which cite articles included in our set of results and were referenced in existing reviews. We perform this step with the aim to gather testing models which are mentioned in other reviews and that were not included in our review due to missing empirical studies. We included the article by Camargo et al.~\cite{camargo_characterising_2015} cited in Rodrigues et al.~\cite{rodrigues_definiton_2010} and manually added 12 articles from the references of the reviews by Afzal et al.~\cite{afzal_software_2016} and Garcia et al.~\cite{garcia_test_2014}. Therefore, the final list of papers consists of 35 papers, 22 included after the application of quality criteria plus 13 added from the other reviews (Fig. \ref{figure:strategy}). Table \ref{table:conferences2} presents the list of conferences/journals for the 13 papers that were added from other reviews.

\begin{table*}[!htb]
\centering
\caption{Conferences/journals ranking of manually added articles.}
\label{table:conferences2}
\begin{footnotesize}
\begin{tabular}{ p{13.6cm} p{1.2cm}} 
 \hline
 \bf{Conference / Journal} & \bf{Rank} \\ 
 \hline 
 International Conference on Software Testing, Verification and Validation Workshops (ICSTW) & A \\
 International Symposium on Empirical Software Engineering (ISESE) & B \\
Journal of Software: Testing, Verification and Reliability & B \\
International Workshop on Software Technology and Engineering Practice (STEP) &	B \\
Journal of Software: Evolution and Process (JSEP) & B \\
Information and Software Technology (I\&ST) & B \\
International Conference on Product-Focused Software Process Improvement (PROFES) &	B \\
Journal of Software Engineering Research and Development (JSERD) & - \\
International Conference on Apperceiving Computing and Intelligence Analysis Proceeding (ICACIA) & - \\
European Conference on Software Process Improvement (EuroSPI) &	- \\
Asia-Pacific Conference on Quality Software (APAQS)	& - \\
International Multitopic Conference (INMIC) & - \\
 \hline
\end{tabular}
\end{footnotesize}
\end{table*}

\subsection{Data extraction \& synthesis} 
Data extraction was performed in two steps to provide answers to the research questions. First, the list of testing models and basic information such as abbreviations and domain were collected from the papers. The models such as MMAST, TAP, TCMM and TOM identified from the article \cite{swinkel_software_2000} were excluded because secondary studies were not included. TPI NEXT, TPI Automotive, CTP and STEP models were also eliminated because we did not obtain any case study. The BS7925-2 standard was replaced by ISO/IEC/IEEE 29119-4, and Test Automation Improvement Model (TAIM) is viewed as work-in-progress and as a research challenge \citep{eldh_towards_2014}, therefore they are not in the final list of testing models. 

For the second step, we extracted more detailed information about the aspects which were improved in the empirical studies, advantages, possible drawbacks and important practices of the models. All this information was synthesized in the SLR results. 

\section{SLR Results}
In this section, we provide the answer to all the SLR research questions (RQ1, RQ2, RQ3, RQ4).

\subsection{Which are the main testing models and their characteristics reported in empirical studies? (RQ1)}\label{section:RQ1}
\noindent Overall, we identified 17 testing models [TM1-TM17] out of the 35 papers (Table \ref{table:approaches}). We describe them in the next sections, as the models will be used to answer all research questions.

\begin{table*}[!htb]
\centering
\caption{Testing Models included in the Systematic Literature Review (\cmark = supported by a maturity model, \xmark = no maturity model)}
\label{table:approaches}
\begin{footnotesize}
\begin{tabular}{ p{0.6cm} p{4.9cm} p{2cm} p{2.6cm} p{1.0cm} p{1.2cm} p{1.4cm} p{1.5cm} } 
 \hline
 \bf{ID} & \bf{Testing Approach} & \bf{Abbreviation} & \bf{Domain} & \bf{Maturity Model} & \bf{Model Base} & \bf{Model Type} & \bf{Reference} \\ 
 \hline
 TM1 & Testing Maturity Model & TMM & General & \cmark & CMM & Model & \cite{burnstein_developing_1996,oh_optimizing_2008} \\
 TM2 & Test Maturity Model integration & TMMi & General & \cmark & TMM & Model & \cite{camargo_identifying_2013,camargo_characterising_2015,rungi_empirical_2013} \\
 TM3 & Test Process Improvement & TPI & General & \cmark & SPICE, TMap & Model & \cite{karlstrom_minimal_2005,ryu_strategic_2008} \\
 TM4 & Embedded Test Process Improvement Model & Emb-TPI & Embedded software & \cmark & TPI & Model &  \cite{jung_test_2009,lee_adapting_2009} \\
 TM5 & Test SPICE & Test SPICE & General & \xmark & ISO 15504 part 5 & Model & \cite{steiner_make_2012} \\
 TM6 & Brazilian Maturity Model for Testing & MPT.BR & Small Organizations & \cmark & ISO/IEC 29119, TMMi & Model & \cite{furtado_mpt.br:_2012} \\
 TM7 & Ministry of National Defense-Testing Maturity Model & MND-TMM & Military systems & \cmark & TMM & Model &  \cite{rodrigues_definiton_2010,ryu_strategic_2008} \\
 TM8 & Software Testing Standard ISO/IEC/IEEE 29119 & ISO/IEC/IEEE 29119 & General & \xmark & ISO & Standard & \cite{ali_formalizing_2015,kasurinen_self-assessment_2011} \\
 TM9 & Test Improvement Model & TIM & General & \cmark & TMM & Model &  \cite{kasurinen_self-assessment_2011} \\
 TM10 & Observing Practice & - & Automation or telecommunication & \xmark & None & Approach & \cite{taipale_improving_2006} \\
 TM11 & Meta-measurement approach & - & General & \xmark & Evaluation theory & Approach & \cite{farooq_meta-measurement_2008} \\
 TM12 & Plan-Do-Check-Action (PDCA) based software testing improvement framework & - & Third party testing center & \xmark & PDCA & Framework & \cite{xu-xiang_pdca-based_2010} \\
 TM13 & Metrics Based Verification and Validation Maturity Model & MB-VV-MM & General & \cmark & TMM & Model & \cite{jacobs_towards_2002} \\
 TM14 & Evidence-based Software Engineering & - & Automotive software & \xmark & None & Approach & \cite{kasoju_analyzing_2013} \\
 TM15 & Self-Assessment framework for ISO/IEC/IEEE 29119 based on TIM & - & General & \cmark & ISO/IEC 29119, TIM & Model &  \cite{kasurinen_self-assessment_2011} \\
 TM16 & Test Process Improvement Model for Automated Test Generation & ATG add-on for TPI & Automated Testing & \cmark & TPI & Model & \cite{heiskanen_test_2012} \\
 TM17 & Minimal Test Practice Framework & MTPF & Small Organizations & \xmark & None & Framework & \cite{karlstrom_minimal_2005} \\
 \hline
\end{tabular}
\end{footnotesize}
\end{table*}

\subsubsection{Testing Maturity Model - TMM}
TMM was developed by the Illinois Institute of Technology (IIT) in 1996 as a complement of CMM because of insufficient improvements of testing processes. The structure of TMM is based on CMM structure, so it defines five maturity levels: Initial, Definition, Integration, Management and Measurement, Optimization/Defect Prevention and Quality Control. Each maturity level has maturity goals and every maturity goal is composed of maturity subgoals. If an organization wants to reach the specific maturity level, they need to accomplish all subgoals within the maturity goals of the level. Except for the maturity model, TMM defines also an assessment model, which contains an assessment procedure, an assessment instrument/questionnaire, team training and selection criteria \citep{oh_optimizing_2008}.

\subsubsection{Test Maturity Model Integration - TMMi}
TMMi was published by the TMMi Foundation in 2007 as the successor of TMM [TM1] and a supplement of the CMMI approach. It consists of five maturity levels: Initial, Managed, Defined, Measured, Optimization; moreover, every level, besides the first level, includes several process areas. The process areas contain generic and specific goals, which can be achieved by the provided practices and their activities \citep{rungi_empirical_2013}. 

\subsubsection{Test Process Improvement - TPI}
TPI [TM3] is a maturity model developed by Martin Poll and Tim Kooman in 1997. It is structured differently than TMM [TM1] or TMMi [TM2]. TPI includes four main elements: Key Areas, Levels, Checkpoints, and Improvement Suggestions. It supports organizations to improve their processes incrementally through 20 key process areas and every key area can have four maturity levels. Checkpoints represent the tool to identify the maturity level of every key area \citep{ryu_strategic_2008}.

\subsubsection{Embedded Test Process Improvement Model - Emb-TPI}
Emb-TPI [TM4] was proposed to consider needs of embedded software and improve strategies for a test capability. It comprises a maturity model, which suggests improvements in different levels, and a capability model to measure the performance for each key area. The next elements of Emb-TPI model are the test evaluation checklist, the evaluation and improvement procedure and the enhanced test evaluation model \citep{jung_test_2009}. 

\subsubsection{Test SPICE}
The Test SPICE model [TM5] was developed with the aim to deliver a process reference model and a process assessment model, which accomplish conformance requirements of ISO/ IEC 15504 II. ISO/IEC 15504 V was used as the starting point with attention to its structure. The original model was transformed by using several methods: transfer a process without any change, replace an original process with a test process, rename process groups and insert a new test process \citep{steiner_make_2012}.

\subsubsection{Brazilian Maturity Model for Testing - MPT.BR}
The MPT.BR model [TM6] is based on ISO-29119-2 [TM8] and TMMi [TM2]. It consists of a reference model, which represents the structure, and an assessment guide, which provides evaluation steps. The reference model defines five maturity levels: Partially Managed, Managed, Defined, Defect Prevention and Automation and Optimization. Each maturity level includes a group of process areas, which provide sets of related practices to achieve a specific goal. The assessment guide is organized into four phases: Hire Assessment, Prepare Assessment, Conduct Assessment and Document Assessment \citep{furtado_mpt.br:_2012}. 

\subsubsection{Ministry of National Defense-Testing Maturity Model - MND-TMM}
MND-TMM [TM7] helps organizations in the defense domain to improve test processes and produce more reliable weapon systems. The structure of the model includes five levels similar to TMM [TM1] and four categories: Military, Process, Infrastructure and Techniques. Each category has several Test Process Areas and they are each further divided into four categories \citep{rodrigues_definiton_2010}.

\subsubsection{Software Testing Standard ISO/IEC/IEEE 29119}
The ISO/IEC/IEEE 29119 standard [TM8] is a set of standards, which support the testing best practices. The model is structured into three layers with the focus on different processes. The first layer is organizational level and includes defined test policies and test strategies within the whole organization. The second layer, called test management, focuses on processes, which define test activities. The last layer includes dynamic test processes to execute tests and create reports \citep{ali_formalizing_2015}.   

\subsubsection{Test Improvement Model - TIM}
The TIM model [TM9] is based on CMM and the goal is to improve test processes from the existing state of the specific organization. The test process consists of five key areas and each key area has five maturity levels: Initial, Baselining, Cost-effectiveness, Risk-lowering and Optimizing \citep{kasurinen_self-assessment_2011}.

\subsubsection{Observing Practice}
The Observing Practice model [TM10] focuses on testing process improvement through a survey of testing practices and interviewing different organizational units. Subsequently, the collected data are processed by using grounded theory \citep{ref:strauss1994grounded}. During this phase, problems and possible solutions are identified \citep{taipale_improving_2006}. 

\subsubsection{Meta-measurement approach}
The Meta-measurement model [TM11] was developed to specify and evaluate quality aspects of the test process. The evaluation process is derived from the Evaluation Theory, which defines six elements: target, evaluation criteria, reference standard, assessment techniques, synthesis techniques and evaluation process. All of these elements were applied in the Meta-measurement approach \citep{farooq_meta-measurement_2008}.

\subsubsection{Plan-Do-Check-Action (PDCA)-based software testing improvement model}
The PDCA-based software testing improvement model [TM12] provides processes to improve service quality and performance for third party testing centers. It focuses mainly on knowledge management (KM) and balance of processes standardization and flexibility. The framework consists of the following phases: building a learning organization through KM, preparing a process-oriented plan,  implementing the plan and collecting data, and continuous improvement \citep{xu-xiang_pdca-based_2010}.

\subsubsection{Metrics Based Verification and Validation Maturity Model - MB-VV-MM}
MB-VV-MM [TM13] is a model, which supports validation and verification processes. It is based on TMM [TM1] with significant improvements. The five-levels structure and most of the process areas are similar to TMM. Compared to TMM, MB-VV-MM includes: comprehensive description of process areas and glossary of terms, new process areas such as V\&V Environment and Organizational Alignment, extended V\&V Training Program process area with career development, Recommended Literature section, metrics for V\&V process improvement and an improved assessment model \citep{jacobs_towards_2002}.

\subsubsection{Evidence-based Software Engineering}
Evidence-based Software Engineering based models [TM14] focus mainly on mapping and gathering evidence through systematic reviews. The process is performed in three steps. First, the need for the evidence and review question are formulated. Subsequently, the evidence is critically evaluated and the answer to the question is provided. In the third step, value stream mapping is used to map the challenges of the testing process to the value stream \citep{kasoju_analyzing_2013}.

\subsubsection{Self-Assessment framework for ISO / IEC / IEEE 29119 based on TIM}
The Self-Assessment framework [TM15] for ISO / IEC / IEEE 29119 based on TIM represents a combination of a maturity level-based approach and a standard-based process. The goal of the model is to provide a simple process assessment tool, which suggests practical objectives on improving the test process. Like ISO/IEC/IEEE 29119 [TM8] standard, the framework processes are divided into three layers: organizational layer, test definition and test execution. Similar to TIM [TM9], the structure consists of five maturity levels: initial, baseline, cost-effectiveness, risk-lowering and optimizing \citep{kasurinen_self-assessment_2011}. 

\subsubsection{Test Process Improvement Model for Automated Test Generation - ATG add-on for TPI}
ATG add-on for TPI model [TM16] reflects particularities of testing based on automated test generation and extends TPI [TM3] by suggesting test improvements for this specific domain. Compared to TPI, ATG add-on for TPI model adds new four key areas: Modeling Approach, Use of Models, Test Confidence, and Technological and Methodological Knowledge. It also modifies some pre-existing key areas in the meaning of new maturity levels and checkpoints \citep{heiskanen_test_2012}. 

\subsubsection{Minimal Test Practice Framework - MTPF}
MTPF [TM17] provides test practices suitable for small and emerging software organizations. The framework consists of five categories and three phases. The categories meet the areas in testing and test organizations: problem and experience reporting, roles and organization issues, verification and validation, test administration, and test planning. The phases are based on the organization growth \citep{karlstrom_minimal_2005}.
\\

Each testing model is specified by a different set of characteristics, domain, and modality of application. The testing models are divided into two big categories based on whether they are supported by a maturity model (Table \ref{table:approaches}): overall, ten out of seventeen testing models have a maturity structure, which defines several improvement stages. The maturity models and their levels differ by goals and key elements used in the testing process. On the other hand, models which are not based on a maturity structure use completely different approaches such as processes divided into three levels [TM8], gathering evidence by performing reviews [TM14] or focusing on knowledge management [TM12].

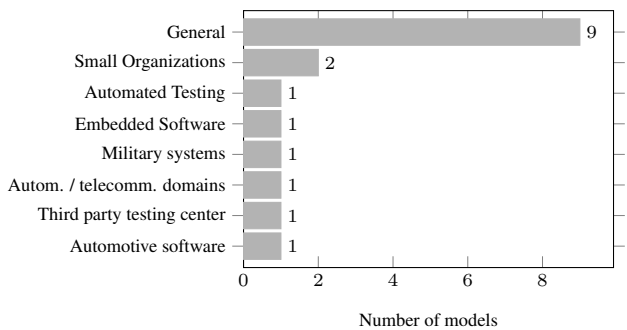
\begin{figure}[!htb]
\centering
\caption{Domains, in which the models are applicable}
\label{figure:domains}
\begin{tikzpicture}
\scriptsize
 \begin{axis}[
 		xbar, xmin=0,
        width=6.5cm, height=5cm, enlarge y limits=0.1,
        xlabel={Number of models},
        symbolic y coords={Automotive software,Third party testing center,Autom. / telecomm. domains,Military systems,Embedded Software,Automated Testing,Small Organizations,General},
 		ytick=data,
        nodes near coords, nodes near coords align={horizontal},
        cycle list = {black!30},
 ]
 		\addplot+[fill,text=black] coordinates {(1,Automotive software) (1,Third party testing center) (1,Autom. / telecomm. domains) (1,Military systems) (1,Embedded Software) (1,Automated Testing) (2,Small Organizations) (9,General)};
 \end{axis}
\end{tikzpicture}
\end{figure}

Based on the empirical studies, we identified the domains in which the models can be applied (Table \ref{table:approaches} \& Fig. \ref{figure:domains}). Nine testing models are defined in literature as universally applicable. On the contrary, several standards have been developed to fulfill the specific needs of the different domains. Because of the broad extent of some general standards, MPT.BR [TM6] and MTPF [TM17] models were designed to satisfy the requirements of small organizations. ATG add-on for TPI [TM16] is suitable for the organizations which pursue automated testing. Emb-TPI [TM4] was developed specifically for testing embedded software, focusing on high quality \citep{jung_test_2009}. For similar reasons, MND-TMM [TM7] is appropriate for military systems. Observing Practice [TM10] was applied in the case study of Taipale and Smolander~\cite{taipale_improving_2006} for automation and telecommunication domains but it is not restricted to these domains. The PDCA-based software testing improvement framework [TM12] was adopted by third party testing centers and Evidence-based Software Engineering model [TM14] was applied for automotive software testing.

\subsection{Which are the reported aspects of a testing process the testing models help to improve? (RQ2)}\label{section:RQ2}
Empirical studies report aspects that the application of testing models helps to improve (Table \ref{table:aspects}). The most reported aspect in the papers is an improvement of the whole testing process by introducing more formalization. Organizations focus particularly on the application of a testing model to improve and standardize testing which can have subsequently positive impact also on other aspects such as product quality \citep{camargo_identifying_2013} or risk management \citep{oh_optimizing_2008}.

Eight case studies focus mainly on quality improvement by adopting a testing model. As it is mentioned in the case study of Farooq et al.~\cite{farooq_meta-measurement_2008}, the quality of a developed product depends mostly on an efficient testing process. For this reason, the authors developed a conceptual framework which defines and evaluates test process quality aspects. When designing a concept of process evaluation, authors were inspired by the Evaluation Theory, which defines six core elements of process evaluation \citep{ref:scriven1991}.						

Another example of the quality improvement is the study of Ryu et al.~\cite{ryu_strategic_2008}, where the authors had to deal with the needs of high quality in the defense domain. They proposed the MND-TMM model [TM7], which can help organizations to implement effective testing and improve testing processes for high quality products. Moreover, they suggested an ontology called MTO (MND-TMM Ontology)  for effective use and execution of model processes, which also can lead to high quality product developments \citep{ryu_strategic_2008}. 

Two papers \citep{eldh_towards_2014,paulish_case_1994} describe and use sets of metrics and measurements as a part of software development improvement. Authors suggest measurements for defects, effort or product size, which can be expressed for example by lines of code (LOC) count. The study of Eldh et al.~\cite{eldh_towards_2014} defines an evaluation process based on validated metrics, which can help to perform test automation in an objective and mature way. In the second case study of Paulish and Carleton~\cite{paulish_case_1994}, the authors measure the effect of methods and design several types of metrics based on the subject of measurement (e.g. defects, effort).

The empirical study of Ali and Yue~\cite{ali_formalizing_2015} designs a conceptual model, which is based on Model-based Testing (MBT) and the ISO/IEC/IEEE 29119 standard, to provide a starting point for developing new MBT techniques.  
In the paper by Kumaresh and Baskaran~\cite{kumaresh_experimental_2012}, authors applied Defect Removal Model to reduce the cost by early removal of the defects. Reid~\cite{reid_bs_2000} focuses on component testing by applying a testing model, which consists of several process phases such as component test planning, specification or execution. The paper of Rodrigues et al.~\cite{rodrigues_definiton_2010} describes how important it is to meet clients' requirements by providing high-quality products. The authors primarily deal with small-sized companies, because not many of them execute test software activities. Authors perform a survey and provide factors, which influence the adoption of a testing process and thus obstruct the final product improvement \citep{rodrigues_definiton_2010}. 

The case study of Xu-Xiang and Zhang~\cite{xu-xiang_pdca-based_2010} applies the PDCA-based testing improvement framework [TM12] based on the analysis of agile processes, benchmarking process and knowledge management. These three areas are parts of the model concept, which address testing application issues and improve the testing service quality. The agile process helps to keep balance between flexibility and stability, and can respond to market changes rapidly. Execution of the benchmarking process improves cooperation, trust and communication, which leads to the creation of an organizational knowledge-oriented culture. Knowledge management includes all the activities that ensure individuals obtain the needed knowledge and subsequently the organization gains competitive advantages \citep{xu-xiang_pdca-based_2010}.

\begin{table}[!htb]
\centering
\caption{Aspects supported by testing models in concrete cases.}
\label{table:aspects}
\begin{footnotesize}
\begin{tabular}{ p{3.2cm} p{4cm} }
 \hline
 \bf{Aspect} & \bf{Publications} \\ 
 \hline 
 Testing Process & \cite{burnstein_developing_1996,camargo_identifying_2013,camargo_characterising_2015,furtado_mpt.br:_2012,heiskanen_test_2012,jacobs_towards_2002,jung_test_2009,karlstrom_minimal_2005,kasoju_analyzing_2013,kasurinen_self-assessment_2011,mirna_advantages_2011,nikula_extending_2009,oh_optimizing_2008,rungi_empirical_2013,steiner_make_2012,swinkel_software_2000} \\
  Quality Improvement & \cite{farooq_meta-measurement_2008,huo_exploratory_2006,lee_adapting_2009,ryu_strategic_2008,simon_spice:_1996,suwanya_analysis_2008,taipale_improving_2006,toroi_identifying_2013} \\
 Metrics / Measurements & \cite{eldh_towards_2014,paulish_case_1994} \\
 Model-based Testing & \cite{ali_formalizing_2015} \\
 Defects Removal and Testing / Inspection Effectiveness & \cite{kumaresh_experimental_2012} \\
 Component Testing & \cite{reid_bs_2000} \\
 Final Product & \cite{rodrigues_definiton_2010} \\
 Agile Processes & \cite{xu-xiang_pdca-based_2010} \\
 Benchmarking & \cite{xu-xiang_pdca-based_2010} \\
 Knowledge Management & \cite{xu-xiang_pdca-based_2010} \\
 \hline
\end{tabular}
\end{footnotesize}
\end{table}


\subsection{Which are the advantages and drawbacks of each testing model as reported in the empirical papers? (RQ3)}\label{section:RQ3}

Based on the article of Cruzes and Dyb\r{a}~\cite{ref:cruzes2010} about synthesizing evidence in research papers, we collect and categorize models' advantages and disadvantages.

\subsubsection{Models' advantages}

There are several advantages that are reported for each of the reviewed models in the empirical studies (Table \ref{table:advantages}). We grouped them into five categories: practice, complementarity, model structure, applicability and other.

\textbf{Practice.} 
As an advantage of two models is considered their practical experience.
The case study of Jacobs and Trienekens~\cite{jacobs_towards_2002} discusses strong points of TMM [TM1] as a baseline of the MB-VV-MM development (TM13): specifically that TMM reflects over forty years of software testing growth in the industry \citep{jacobs_towards_2002}. 
Based on the study of Ryu et al.~\cite{ryu_strategic_2008}, the TPI model [TM3] provides guidelines for maturity level assessment, which are inspired by practical experience of the test process in organizations.

\textbf{Complementarity.}
According to the study of Jacobs and Trienekens~\cite{jacobs_towards_2002}, the TMM model can be used to complement CMM, which does not sufficiently deal with testing issues \citep{jacobs_towards_2002}. 
Rungi and Matulevičius~\cite{rungi_empirical_2013} analyze the TMMi model [TM2]  and its benefits. They consider TMMi as a complementary model to CMMI: experience and results of the TMMi assessment procedure can be used to perform any future evaluation according to CMMI \citep{rungi_empirical_2013}.

\textbf{Model Structure.}
Several advantages come from the models' structure.
The research of Rungi and Matulevičius~\cite{rungi_empirical_2013} also discusses the structure of the TMMi model [TM2]. The evolutionary staged model of TMMi maturity levels allows evaluation of the current status and provides a clear improvement path for achievement of a higher maturity level \citep{rungi_empirical_2013}.
Furthermore, Heiskanen et al. \cite{heiskanen_test_2012} recognize benefits of TPI for the ATG add-on for TPI [TM16] development. One of the strengths of TPI that the authors report is the possibility to progress simultaneously in many different Key Areas, which allows organizations to focus on more selected areas \citep{heiskanen_test_2012}.
The significant characteristic of the TIM [TM9] model discussed in the case study of Kasurinen et al.~\cite{kasurinen_self-assessment_2011} is a separate assessment and independence of the key areas. Organizations can better understand what areas need to be improved \citep{kasurinen_self-assessment_2011}. 
The study of Farooq et al.~\cite{farooq_meta-measurement_2008} focuses on a theoretical and conceptual representation of the Meta-measurement approach [TM11]. It evaluates the model as a lightweight framework, with the strength of the independence from the development life cycle as well as from any testing process in use.
One of the strengths of MTPF [TM17] discussed in the study of Karlström et al.~\cite{karlstrom_minimal_2005} is the specification of small steps, which allow to follow an easier improvement path. 

\textbf{Applicability.}
Model applicability is one of the key points in the selecting process.
According to the study of Heiskanen et al.\cite{heiskanen_test_2012}, TPI is suitable for specific test projects or approaches thanks to its natural technical base \citep{heiskanen_test_2012}.
The empirical study of Ali and Yue~\cite{ali_formalizing_2015} focuses on a conceptual model based on ISO/IEC/IEEE 29119 [TM8]. The authors consider the standard defined at a very high level of abstraction, that can help to apply it to different types of testing techniques. For that reason, they use a small subset of the standard as a conceptual model to understand terminology and define key concepts of the Model-based Testing technique.

\textbf{Other.}
Two advantages cannot be categorized because of their specialization. 
Xu-Xiang and Zhang~\cite{xu-xiang_pdca-based_2010} have developed the PDCA-based software testing improvement framework [TM12] to address testing application issues. Based on the analysis, the framework is shown to be flexible, easy-to-use and helps to influence the initiative of the staff. Moreover, knowledge management testing activities support the reuse of testing cases and help to raise the testing center’s competitive advantage \citep{xu-xiang_pdca-based_2010}.
In comparison to other models, MTPF introduces each phase using the low threshold and ensures the validity by using rigorous and well-established research methodologies \citep{karlstrom_minimal_2005}.

\begin{table*}[!htb]
\centering
\caption{Advantages of testing models}
\label{table:advantages}
\begin{footnotesize}
\begin{tabular}{ p{3cm} p{14cm} } 
 \hline
 \bf{Category} & \bf{[Model ID:] Advantages} \\
 \hline 
 Practice & \begin{itemize}[leftmargin=*]
 \item [TM1:] Many years of \textbf{industrial experience} with software testing \citep{jacobs_towards_2002}.
 \item [TM3:] Developed based on the \textbf{practical knowledge} and experiences \citep{ryu_strategic_2008}. 
 \end{itemize} \\
 Complementarity& \begin{itemize}[leftmargin=*]
 \item [TM1:] A \textbf{counterpart of} the \textbf{CMM} model \citep{jacobs_towards_2002}.
 \item [TM2:] A \textbf{complement model to CMMI}; helps to perform any evaluations according to CMMI \citep{rungi_empirical_2013}.
 \end{itemize} \\
 Model Structure& \begin{itemize}[leftmargin=*]
  \item [TM2:] The evolutionary staged model allows \textbf{assessment of the current state} of maturity and suggests \textbf{a clear improvement path} \citep{rungi_empirical_2013}.
  \item [TM3:] It allows to \textbf{achieve} different \textbf{Key Areas simultaneously} \citep{heiskanen_test_2012}.
   \item [TM9:] \textbf{Key Areas} are \textbf{assessed separately} and are relatively \textbf{independent} \citep{kasurinen_self-assessment_2011}.
   \item [TM11:] \textbf{Independent from the development life cycle} and any testing process being followed \citep{farooq_meta-measurement_2008}.
   \item [TM17:] The \textbf{small steps} are intended to provide an easier path to follow \citep{karlstrom_minimal_2005}.
 \end{itemize} \\
 Applicability& \begin{itemize}[leftmargin=*]
 \item [TM3:] Appropriate for specific test projects because it is naturally \textbf{technical} \citep{heiskanen_test_2012}.
 \item [TM8:] Can be \textbf{applied to different} types of testing \textbf{techniques} \citep{ali_formalizing_2015}. 
 \end{itemize} \\
 Other& \begin{itemize}[leftmargin=*]
\item [TM12:] It focuses also on \textbf{individual’s subjective initiative} and competitive advantage \citep{xu-xiang_pdca-based_2010}.
\item [TM17:] Compared to other models, MTPF provides the \textbf{low threshold} for introducing each phase \citep{karlstrom_minimal_2005}.
 \end{itemize} \\
 \hline
\end{tabular}
\end{footnotesize}
\end{table*}

\subsubsection{Models' drawbacks}

Empirical papers also report impediments and critical factors in the adoption of testing models (Table \ref{table:drawbacks}). We grouped them into the following categories: lack of applicability, lack of support for specific areas, model complexity, practical limitations, other aspects.

\textbf{Lack of applicability.}
One of the models' disadvantages is the lack of applicability in some cases.
Testing requirements of smaller organizations are discussed in the study of Karlström et al.~\cite{karlstrom_minimal_2005}. The paper focuses on adaptation of the MTPF [TM17] framework for small- to medium-sized enterprises (SME). In the first part, the authors review existing testing models (e.g., TMM [TM1], TPI [TM3], TIM [TM9]) and their usability for SME. Although the models are applicable in general, organizations usually consider that models are not suitable for SME and furthermore insufficient for specific domains such as defense or embedded software implementation, because they are too extensive. Moreover, the resources needed for the adaptation of the models are not available in smaller organizations. In that case, SMEs would need to make a lot of effort to establish model's processes \citep{karlstrom_minimal_2005}. 

The empirical study reviewing TMMi through a survey published by Camargo et al.~\cite{camargo_characterising_2015}, extends the results presented in \cite{camargo_identifying_2013}. The authors observed that TMMi and MPT.BR have too many practices which have to be satisfied and not all of them are applicable for smaller-sized companies. Furthermore, they do not define priorities in case of absence of time or resources. 

\textbf{Lack of support for specific areas.}
Prioritization plays a major role in software engineering \cite{ref:pergher2013requirements}. Several specific areas such as prioritizing or test automation are not supported by some models.
The case study of Jung~\cite{jung_test_2009} performed analysis of existing testing models with the aim to find a suitable model for embedded software. The authors discuss disadvantages of the TMM [TM1] and TPI [TM3] models and provide reasons why they are insufficient for embedded software which requires high quality. Based on their findings, test elementary factors are not taken into account in TMM, specifically test organization, human resource management and test bed. Moreover, TMM does not describe procedures and guidelines for process evaluation and improvement in detail \citep{jung_test_2009}. 

The TPI model also does not meet the embedded software requirements, because it does not provide practical and realistic improvements on that kind of software \citep{jung_test_2009}. 

The weaknesses of TMM [TM1] are mentioned also in the study of Oh et al.~\cite{oh_optimizing_2008}, which highlights the issue of prioritizing multiple actions. The paper deals with an effective implementation strategy for the models and it includes also ranking importance of improvements activities. However, TMM does not provide any procedure for prioritizing multiple actions which focus on the same maturity goals. This drawback can lead to failure of the activities and wasting the organizational resources without any positive effects \citep{oh_optimizing_2008}. 

The case study of Ryu et al.~\cite{ryu_strategic_2008} includes an initial part about reviewing the testing models (e.g., TMM [TM1], TPI [TM3], TIM [TM9]), their characteristics and limitations. First, the study describes TMM and its usage as a supplementary model of the CMM. According to this relation, it is more effective to apply them together, while based on the study it is difficult to use only the TMM or even combine with other models \citep{ryu_strategic_2008}. Other drawbacks of TMM are lack of important key areas such as Test Environment, Office Environment, Reporting, Defect Management, and Testware Management and high complexity of included key areas, that makes them difficult to achieve \citep{ryu_strategic_2008}. 

The study of Camargo et al.~\cite{camargo_identifying_2013} presents a survey performed among software testing professionals. The paper reports several limitations of TMMi [TM2] based on the results from the review. According to the review the model does not define any priorities of activities or dependencies between the process areas \citep{camargo_identifying_2013}. 

Eldh et al.~\cite{eldh_towards_2014} comment in their study that although TMMi is adapted for new processes (e.g. agile), it does not provide improvement steps for real test automation advancement. The study of Rungi and Matulevičius~\cite{rungi_empirical_2013}, which mentions TMMi weaknesses, indicates that the model does not include detailed documentation about agile practices in comparison with CMMI and the application of best practices.

\textbf{Model complexity.}
About the structure and features of the TPI model, the authors consider incompatibility of TPI with the CMM / CMMI and TMM caused by different structure of levels, and missing checking of items for state-of-the-art testing techniques after their publishing \citep{ryu_strategic_2008}. In general, these models are too complex and difficult for organizations to apply and because of many procedures and process areas they require the process assessment from experts to inspect the gaps \citep{ryu_strategic_2008}.

The study of Camargo et al.~\cite{camargo_identifying_2013} considers the TMMi model structure as very comprehensive, therefore it can be confusing and demanding to understand differences between some activities.

The study of Ali and Yue~\cite{ali_formalizing_2015} uses ISO/IEC/IEEE 29119 [TM8] as a conceptual model for Model-Based Testing techniques. Based on their experience, the model has very high abstraction level, that can lead to misunderstandings about the concepts. The authors solve this issue by developing the conceptual model and subsequently demonstrating it in the context of different types of testing techniques.

\textbf{Practical limitations.}
Based on the experience with MTPF implementation within an organization and a performed survey in other companies, the authors conclude the model as applicable in SMEs but requiring continuous adaptation of the practices in order for them to stay current, effective and useful \citep{karlstrom_minimal_2005}.  

At the beginning, the study of Kasurinen et al.~\cite{kasurinen_self-assessment_2011} discusses practical limitations of TMMi application in real-life companies. Specifically, the idea of moving all processes to one level before achieving the next one is not feasible, and the level requirements are counter-intuitive.

\textbf{Others.}
The PDCA-based software testing improvement framework [TM12] is developed and published by Xu-Xiang and Zhang~\cite{xu-xiang_pdca-based_2010}. The framework is designed to address the issues of testing application and knowledge management (KM) implementation by focusing mostly on agile process, benchmarking process and KM. Although the framework is flexible and efficient, the authors defined several constraints which can influence the effect. First, organizations should construct and support the knowledge-oriented culture and sustain the enthusiasm from the teams. Second, during planning and adaptation of the testing activities, organizations should understand popular testing models completely. And third, KM has to be included at the organizational level and managers within the organizations should support it \citep{xu-xiang_pdca-based_2010}.

Although the Self-Assessment framework for ISO / IEC / IEEE 29119 based on TIM [TM15] suggested by Kasurinen et al.~\cite{kasurinen_self-assessment_2011}  can become a feasible tool for defining process improvement objectives, it needs further development to be validated as a testing model for real cases.

\begin{table*}[!htb]
\centering
\caption{Drawbacks of testing models}
\label{table:drawbacks}
\begin{footnotesize}
\begin{tabular}{ p{4.5cm} p{12cm} } 
 \hline
 \bf{Category} & \bf{[Model ID:] Drawbacks} \\
 \hline 
 Lack of applicability & \begin{itemize}[leftmargin=*]
  \item [TM1,TM3,TM9:] \textbf{Not} applicable \textbf{for small- to medium-sized enterprises} (SME) \citep{karlstrom_minimal_2005}.
   \item [TM1,TM3,TM9:] Several \textbf{limitations} in the implementation \textbf{for specific domain}. \textbf{Too complex} and difficult to adapt for a specific case, requires the experts’ process evaluation, difficult to apply without CMM or combine with other models and lack of important key areas \citep{ryu_strategic_2008}.
 \end{itemize} \\
 Lack of support for specific areas & \begin{itemize}[leftmargin=*]
  \item [TM1:] \textbf{No support for test basic elementary factors} (e.g. test organization, human resource management and test bed \citep{jung_test_2009}.
 \item [TM1:] \textbf{No procedures for prioritizing} of multiple actions focused on the same maturity goals \citep{oh_optimizing_2008}.
 \item [TM2:] \textbf{Priorities are not defined} in the case of lack of time or resources \citep{camargo_characterising_2015}.
 \item [TM2:] \textbf{The description} of improvement steps \textbf{for test automation is limited} \citep{eldh_towards_2014}.
  \item [TM2:] \textbf{No comprehensive documentation about agile practices} in comparison with CMMI and application of best practices \citep{rungi_empirical_2013}.
  \item [TM3:] \textbf{No practical} and realistic \textbf{improvements on embedded software} \citep{jung_test_2009}.
  \item [TM6:] \textbf{Priorities are not defined} in the case of lack of time or resources \citep{camargo_characterising_2015}. 
  \end{itemize} \\
 Model complexity & \begin{itemize}[leftmargin=*]
   \item [TM2:] Very detailed practices and distribution across the specific goals and process areas make the model \textbf{complex and difficult} to understand \citep{camargo_identifying_2013}.  
   \item [TM8:] \textbf{A high level of abstraction} introduces ambiguity in understanding the concepts \citep{ali_formalizing_2015}.
 \end{itemize} \\
 Practical limitations & \begin{itemize}[leftmargin=*]
  \item [TM2:] \textbf{Practical limitations} in real cases, for example the implementation of the idea to move all processes to one level before proceeding to the next is not possible and \textbf{the level requirements are not intuitive} \citep{kasurinen_self-assessment_2011}.
   \item [TM17:] It \textbf{requires continuous adaptation} of the practices \citep{karlstrom_minimal_2005}. 
 \end{itemize} \\
 Other & \begin{itemize}[leftmargin=*]
 \item [TM12:] The organization \textbf{should construct} and strengthen \textbf{the knowledge-oriented culture} and the team enthusiasm.
 \item [TM12:] The organization \textbf{should understand the model} fully when planning and adapting the testing processes.
 \item [TM12:] \textbf{The knowledge management} strategy needs to be included at the organizational level \citep{xu-xiang_pdca-based_2010}. 
 \item [TM15:] The model \textbf{needs further development} and studies for validity \citep{kasurinen_self-assessment_2011} 
 \end{itemize} \\
 \hline
\end{tabular}
\end{footnotesize}
\end{table*}

\subsection{Which are the phases/practices of testing models for different domains? (RQ4)}\label{section:RQ4}
Models also define practices which are important for a testing process (Tables \ref{table:practicesPlanning}, \ref{table:practicesDesign}, \ref{table:practicesSetup}, \ref{table:practicesExecute} and \ref{table:practicesMonitor}). We collect the practices based on models' phases ( \textit{i.~Planning}, \textit{ii.~Test Case Design}, \textit{iii.~Setup of test environment and data}, \textit{iv.~Execution and Evaluation}, and \textit{v.~Monitoring and Control}) and map them to specific domains. The embedded software domain is not included in the tables, because no paper discusses the practices of the Emb-TPI model. 

\textbf{General.}
We collect practices of three models.
The research of Camargo et al.~\cite{camargo_identifying_2013} identifies practices of the TMMi model [TM2] based on a survey. The authors define three profiles to analyze the results regarding the level of importance. They created a reduced set of practices (31 items), which are important for all three profiles \cite{camargo_identifying_2013}. The first phase is \textit{Planning}, which defines how testing will be executed and what will be tested. The Planning phase includes practices such as \textit{Identify product risks} and \textit{Analyze product risks}, with outputs necessary for test prioritization. Other practices \textit{Identify items and features to be tested, Establish the test schedule} and \textit{Plan for test staffing} are also marked as mandatory because they are included in the test plan and thus related to the practice \textit{Establish the test plan}. The phase \textit{Test Case Design} include only two mandatory practices: \textit{Identify and prioritize test cases} and \textit{Identify necessary specific test data}. During the \textit{Setup of Test Environment and Data} phase the tests and requirements are prioritized and implemented. The order is defined based on the product risks. There are practices such as \textit{Develop and prioritize test procedures, Develop test execution schedule, Implement the test environment} and \textit{Perform test environment intake test}. During the \textit{Execution and Evaluation} phase test are performed and possible defects are reported. Moreover, results and the achievement of test goals are evaluated. Status checking and, if necessary, corrective actions are parts of the \textit{Monitoring and Control} phase \cite{camargo_identifying_2013}.

The study of Farooq et al.~\cite{farooq_meta-measurement_2008} describes practices of the Meta-measurement approach [TM11]. The first step includes establishing evaluation requirements which represent the evaluator's needs and objectives. The second step is specifying evaluation scope and measurements. Then, procedures for the evaluation activities are defined and as the last step the evaluation is executed and measurements are collected \cite{farooq_meta-measurement_2008}.

Jacobs and Trienekens~\cite{jacobs_towards_2002} discuss generic practices of the MB-VV-MM model [TM13] and group them into several common features. The first group is \textit{Commitment to Perform}, which includes a practice for defining organizational expectations for the process. The second feature \textit{Ability to Perform} ensures availability of needed resources and assigning the responsibility for performing the process and achieving results. Furthermore, training is performed to ensure the necessary skills to execute the process. The feature \textit{Activities Performed} describes the activities that have to be executed to establish the process. Configuration and process measurements are managed within the \textit{Directing Implementation} feature. The process is monitored and controlled to collect information and if necessary, perform corrective actions. The last feature \textit{Verifying Implementation} evaluates if the process is implemented as it was planned and provides business management \cite{jacobs_towards_2002}.

\begin{table*}[!htb]
\centering
\caption{Mapping models' practices and domains - Planning phase (\xmark = practice applicable)}
\label{table:practicesPlanning}
\begin{tabular}{ | p{6.6cm} | c | c | c | c | c | c | c |} 
 \hline
 \diagbox[width=6cm]{\bf{Planning Practices}}{\bf{Domain}} & \begin{turn}{90}\bf{General} \end{turn}& \begin{turn}{90}\bf{Small Org.}\end{turn} & \begin{turn}{90}\bf{Automated Testing}\end{turn} & \begin{turn}{90}\bf{Military Systems}\end{turn} & \begin{turn}{90}\bf{Autom./Tele\-com.}\end{turn} & \begin{turn}{90}\bf{Testing center}\end{turn} & \begin{turn}{90}\bf{Automotive sw}\end{turn} \\
 \hline 
Involve risk management & \xmark & \xmark & & & & & \\ \hline
Identify items and features to be tested & \xmark & & & & & & \\ \hline
Define the test approach & \xmark & & & & & & \\ \hline
Define exit criteria & \xmark & & & & & & \\ \hline
Establish the test schedule & \xmark & & & & & & \\ \hline
Plan for test staffing & \xmark & & & & & & \\ \hline
Establish the test plan & \xmark & \xmark & \xmark & \xmark & \xmark & \xmark & \xmark \\ \hline
Elicit test environment needs & \xmark & & & & & & \\ \hline
Analyze the test environment requirements & \xmark & & & & & \xmark & \xmark \\ \hline
Establish Evaluation Requirements & \xmark & & & & & & \\ \hline
Specify Evaluation & \xmark & & & & & & \\ \hline
Define Evaluation Scope & \xmark & & & & & & \\ \hline
Define required measurements & \xmark & & & & & &\\ \hline
Define reference standards & \xmark & & & & & & \\ \hline
Establish an organizational policy & \xmark & \xmark & & & & & \\ \hline
Assign responsibility & \xmark & \xmark & & & & &\\ \hline
Train people & \xmark & \xmark & & & & & \\ \hline
Cover non-functional requirements & & & \xmark & & & & \\ \hline
Identify Software Process Life Cycle & & & & \xmark & & & \\ \hline
Select case organizational units & & & & & \xmark & & \\ \hline
Plan agile testing process & & & & & & \xmark & \\ \hline
\end{tabular}
\end{table*}

\textbf{Small Organizations.}
The practices of MPT.BR [TM6] for small organizations are discussed in the study of Furtado et al.~\cite{furtado_mpt.br:_2012}. It includes generic practices for the first two maturity levels (Partially managed and Managed). Other three levels do not add extra generic practices. A generic practice reflects process capabilities that are important for process areas of maturity levels \cite{furtado_mpt.br:_2012}. 

Another model designed for small organizations is MTPF [TM17], which practices are described in the study of Karlstr\"{o}m et al.~\cite{karlstrom_minimal_2005}. The model defines several categories and related phases. At the beginning test planning is performed with the aim to create a test plan, coordinate software quality assurance, involve risk management and define responsibilities and roles. During the second phase, test cases are derived to ensure that the most common actions will be tested. Subsequently, the test environment has to be organized and available for testing when needed. At the end, tests are performed and the test process is evaluated~\cite{karlstrom_minimal_2005}.

\textbf{Automated Testing.}
The Automated Testing domain is specific for the ATG add-on for TPI model [TM16] and it is discussed in the research of Heiskanen et al.~\cite{heiskanen_test_2012}. The practices are defined separately for all maturity levels and key areas. We collect and map them into our phases. During the planning phase, the project is initialized and a plan is created. Then, it is important to design and model test basis and subsequently to conduct techniques and environment. After performing tests, the strategy is evaluated and optimal test automation is conducted. The monitoring and control phase includes conducting organization metrics and risk and defect management~\cite{heiskanen_test_2012}.

\begin{table*}[!htb]
\centering
\caption{Mapping models' practices and domains - Test Case Design phase (\xmark = practice applicable)}
\label{table:practicesDesign}
\begin{tabular}{ | p{8.5cm} | c | c | c | c | c | c | c |} 
 \hline
 \diagbox[width=7.5cm]{\bf{Design Practices}}{\bf{Domain}} & \begin{turn}{90}\bf{General} \end{turn}& \begin{turn}{90}\bf{Small Org.}\end{turn} & \begin{turn}{90}\bf{Automated Testing}\end{turn} & \begin{turn}{90}\bf{Military Systems}\end{turn} & \begin{turn}{90}\bf{Autom./Tele\-com.}\end{turn} & \begin{turn}{90}\bf{Testing center}\end{turn} & \begin{turn}{90}\bf{Automotive sw}\end{turn} \\
 \hline 
Identify and design test cases & \xmark & \xmark & \xmark & \xmark & & & \xmark \\ \hline
Prioritize test cases & \xmark & & & & & & \\ \hline
Identify necessary specific test data & \xmark & & & & & & \\ \hline
Design evaluation & \xmark & & & & & & \\ \hline
Provide resources & \xmark & & & & & & \\ \hline
Provide senior management with visibility of the process & & \xmark & & & & & \\ \hline
Collect data & & & & & \xmark & \xmark & \\ \hline
Implement agile testing process definition & & & & & & \xmark & \\ \hline
Identify and design test scripts & & & & & & & \xmark \\ \hline
\end{tabular}
\end{table*}

\textbf{Military Systems.}
The MND-TMM model [TM7] is designed and developed for military systems. The study of Ryu et al.~\cite{ryu_strategic_2008} provides practices only for the first phase. The goal of the planning phase is to set a test plan which describes the activities needed at a specific level, and identify software development life cycle performed at each level~\cite{ryu_strategic_2008}.  

\textbf{Automation or Telecommunication Domains.}
The Observing Practice [TM10] was applied in the Automation and telecommunication domain. Its practices are presented in the reseach of Taipale and Smolander~\cite{taipale_improving_2006}. It is a very specific approach, which is based on collecting data through interviews. From that reason, the practices represent steps performed during the interviews. The first step includes selecting case organizational units, e.g. the study focuses on the units that develop and test technical software for automation or telecommunication domains in Finland. As a next step, data are collected by conducting interviews with respondents to elicit views and experiences. Subsequently, data are analyzed by using the grounded theory and based on the results process improvements are proposed and applied \cite{taipale_improving_2006}. 

\textbf{Third party testing center.}
The third party testing center domain is presented in the PDCA-based software testing improvement framework [TM12]. The study of Xu-Xiang and Wen-Ning~\cite{xu-xiang_pdca-based_2010} describes the framework process in several phases. The process starts with building the learning organization and requirements analysis. The process-oriented plan is prepared which includes adaptive testing processes based on the organizational processes. Then, it is possible to implement the plan and collect data, which are analyzed with the aim to identify process improvements. When improvement actions are prepared they can be implemented and the results are assessed \cite{xu-xiang_pdca-based_2010}. 

\textbf{Automotive software.}
The study of Kasoju et al.~\cite{kasoju_analyzing_2013} discusses the practices of the Evidence-based Software Engineering approach [TM14] specific for automotive software. The authors define test process activities based on the analysis of collected data. The planning phase includes estimation of the requirements, test techniques, tools and other test artifacts, and creation of a test plan. During the analysis and design phase, test scripts and test data are identified and designed to determine how the tests will be performed. Then, required test environment and other important artifacts have to be collected and built. When everything is prepared, tests can be executed and reports of results can be send to the customer \cite{kasoju_analyzing_2013}.

\begin{table*}[!htb]
\centering
\caption{Mapping models' practices and domains - Setup of test environment and data phase (\xmark = practice applicable)}
\label{table:practicesSetup}
\begin{tabular}{ | p{8.9cm} | c | c | c | c | c | c | c |} 
 \hline
 \diagbox[width=7.5cm]{\bf{Setup Practices}}{\bf{Domain}} & \begin{turn}{90}\bf{General} \end{turn}& \begin{turn}{90}\bf{Small Org.}\end{turn} & \begin{turn}{90}\bf{Automated Testing}\end{turn} & \begin{turn}{90}\bf{Military Systems}\end{turn} & \begin{turn}{90}\bf{Autom./Tele\-com.}\end{turn} & \begin{turn}{90}\bf{Testing center}\end{turn} & \begin{turn}{90}\bf{Automotive sw}\end{turn} \\
 \hline 
Develop test procedures & \xmark & \xmark & \xmark & \xmark & & & \xmark \\ \hline
Prioritize test procedures & \xmark & & & & & & \\ \hline
Develop test execution schedule & \xmark & & & & & & \\ \hline
Implement the test environment & \xmark & \xmark & \xmark & & & & \xmark \\ \hline
Perform test environment intake test & \xmark & & & & & & \\ \hline
Develop evaluation & \xmark & & & & & & \\ \hline
Manage configurations & \xmark & & \xmark & & & & \\ \hline
Identify and provide resources & & \xmark & & & & & \\ \hline
Analyze data & & & & & \xmark & & \\ \hline
Conduct benchmarking and interlaboratory comparison & & & & & & \xmark & \\ \hline
Determine the optimum improvement items & & & & & & \xmark & \\ \hline
Establish the improvement objectives & & & & & \xmark & \xmark & \\ \hline
\end{tabular}
\end{table*}

\begin{table*}[!htb]
\centering
\caption{Mapping models' practices and domains - Execution and Evaluation phase (\xmark = practice applicable)}
\label{table:practicesExecute}
\begin{tabular}{ | p{8.9cm} | c | c | c | c | c | c | c |} 
 \hline
 \diagbox[width=8.5cm]{\bf{Execution Practices}}{\bf{Domain}} & \begin{turn}{90}\bf{General} \end{turn}& \begin{turn}{90}\bf{Small Org.}\end{turn} & \begin{turn}{90}\bf{Automated Testing}\end{turn} & \begin{turn}{90}\bf{Military Systems}\end{turn} & \begin{turn}{90}\bf{Autom./Tele\-com.}\end{turn} & \begin{turn}{90}\bf{Testing center}\end{turn} & \begin{turn}{90}\bf{Automotive sw}\end{turn} \\
 \hline 
Execute test cases & \xmark & \xmark & \xmark & \xmark & & & \xmark \\ \hline
Report test incidents & \xmark & & & & & & \\ \hline
Write test log & \xmark & & & & & & \\ \hline
Decide disposition of test incidents in configuration control board & \xmark & & & & & & \\ \hline
Perform appropriate action to close the test incident & \xmark & & & & & & \\ \hline
Track the status of test incidents & \xmark & & & & & & \\ \hline
Execute non-functional test cases & \xmark & & & & & & \\ \hline
Report non-functional test incidents & \xmark & & & & & & \\ \hline
Write test log & \xmark & & & & & & \\ \hline
Execute Evaluation & \xmark & \xmark & \xmark & & & & \xmark \\ \hline
Collect selected measurements & \xmark & & & & & & \\ \hline
Conduct optimal test automation & & & \xmark & & & & \\ \hline
Apply process improvements & & & & & \xmark & \xmark & \\ \hline
\end{tabular}
\end{table*}

\begin{table*}[!htb]
\centering
\caption{Mapping models' practices and domains - Monitoring and Control phase (\xmark = practice applicable)}
\label{table:practicesMonitor}
\begin{tabular}{ | p{8cm} | c | c | c | c | c | c | c |} 
 \hline
 \diagbox[width=8.3cm]{\bf{Monitoring Practices}}{\bf{Domain}} & \begin{turn}{90}\bf{General} \end{turn}& \begin{turn}{90}\bf{Small Org.}\end{turn} & \begin{turn}{90}\bf{Automated Testing}\end{turn} & \begin{turn}{90}\bf{Military Systems}\end{turn} & \begin{turn}{90}\bf{Autom./Tele\-com.}\end{turn} & \begin{turn}{90}\bf{Testing center}\end{turn} & \begin{turn}{90}\bf{Automotive sw}\end{turn} \\
 \hline 
Conduct test progress reviews & \xmark & & & & & & \\ \hline
Monitor defects & \xmark & & \xmark & & & & \xmark \\ \hline
Record defects & & & & & & & \xmark \\ \hline
Conduct product quality reviews & \xmark & & & & & & \\ \hline
Analyze issues & \xmark & & & & & & \\ \hline
Take corrective action & \xmark & & & & & & \\ \hline
Manage corrective action & \xmark & & & & & & \\ \hline
Co-ordinate the availability and usage of the test environments & \xmark & & & & & & \\ \hline
Report and manage test environment incidents & \xmark & & & & & & \xmark \\ \hline
Evaluate results & \xmark & & & & & & \\ \hline 
Suggest improvements & \xmark & & & & & & \\ \hline
Measure the process results & \xmark & &\xmark & & & & \\ \hline
Review status with business management & \xmark & & & & & & \\ \hline
Monitor and control the process & & \xmark & & \xmark & & & \\ \hline
Control work products & & \xmark & & & & & \\ \hline
Assess the improvement result & & & & & & \xmark & \\ \hline
Manage risks & & &\xmark & & & & \\ \hline
\end{tabular}
\end{table*}

\section{Discussion}
The SLR on software testing models was conducted to identify existing testing models and their characteristics such as domain, context of application, aspects improved, advantages and drawbacks, which can help organizations in selecting an appropriate model for a testing process. 

The first question [RQ1] focused on the list of existing testing models gathered from the empirical studies presented in Section \ref{section:RQ1} and models characteristics such as the domain of application. 

The final list of models is similar to the second survey of Afzal et al.~\cite{afzal_software_2016} except TPI-Automotive and TPI NEXT. Moreover, we included the MPT.BR model [TM6], which was developed for small organizations and is applied by several software organizations in Brazil \citep{furtado_mpt.br:_2012}.
Because of insufficient set of case studies, we manually added several articles referenced in the survey of Afzal et al.~\cite{afzal_software_2016}, which lead to extended information. Overall we included seventeen models, with ten that are supported by a maturity model.

The second part of the first question relates to the expanded information about the models domain, for which the models are applicable. This knowledge is based on the area in which the empirical studies or the model application were performed. The results are shown in Section \ref{section:RQ1}. Several standards were developed specifically for one domain such as Emb-TPI [TM4] for embedded software or ATG add-on for TPI [TM16] for Automated testing. On the other hand, some models were applied to a specific domain, but they do not have to be limited only to this area: for instance the PDCA-based software testing improvement framework [TM12] was applied to third party testing centers \citep{xu-xiang_pdca-based_2010}.

The aspects that were improved by the testing models are discussed in the second question [RQ2] and answered in Section \ref{section:RQ2}. We observed that the most supported aspect in the studies is the improvement of the overall testing process by means of more formalization. Through the application of models, organizations improved and standardized their way of testing, which significantly influenced the product quality. Some studies focus more on other aspects such as measurements, defects or agile processes.

The third question [RQ3] focuses on the advantages and drawbacks of approaches in concrete cases. Tables of summarized information are presented in Section \ref{section:RQ3}. The authors of empirical studies usually interpreted the drawbacks of testing models as a reason to develop new testing approaches, which will cover the deficiencies. On the other hand, the benefits came from the successful implementation of the models. Compared to other surveys, Garcia et al.~\cite{garcia_test_2014} summarize details about models but they are not very extensive. 

The last question [RQ4] includes the mapping of models' phases and practices to the specific domains. The results are presented in Section \ref{section:RQ4}. The practices of models, which are not developed for a specific domain are better documented and presented in the studies, specifically the TMMi model. The Embedded Software domain is not included in the results because there are no practices reported in the study. Practices for the Military Systems domain are discussed only for the planning phase, but most of the practices can be taken from the general models.

\section{Threats to validity}
Evidence Based Software Engineering (EBSE) emerged over the years as a way to support software engineering practices with a more scientific approach  \citep{ref:kitchenham2004evidence,ref:dyba2005evidence}. Since the emergence of EBSE, a large number of SLRs were performed in the area \citep{ref:kitchenham2009systematic}. However, running SLRs poses several challenges in terms of limitations and validity threats, being many of the phases subjective with possibilities of introducing researchers bias and errors \citep{ref:petersen2015guidelines,ref:budgen2006performing,ref:kitchenham2004procedures}. 

There are many classifications of validity threats that can be applied to software engineering empirical studies (e.g. for case studies -- \cite{ref:runeson2009guidelines}, or experiments in software engineering -- \cite{ref:wohlin2012experimentation}). To report threats to validity, we selected the guidelines proposed by Petersen and Gencel~\cite{ref:petersen2013worldviews} and used in \cite{ref:petersen2015guidelines}, as they are particularly fit for qualitative studies such as SLRs---slightly different than traditional  validity threats (\textit{internal}, \textit{construct}, \textit{conclusion}, \textit{external} \textit{threats}) typically used in software engineering studies \citep{ref:wohlin2012experimentation}. 

\textbf{Theoretical Validity.} Theoretical validity refers to threats in building theory out of the observed phenome\-non~\citep{ref:petersen2013worldviews}.

Researcher bias is a theoretical threat in the selection of studies that were part of the SLR. The SLR was conducted by one main researcher, however, the selection of articles was validated by a second researcher to reduce this selection bias. Final results of included papers might be influenced by the views of the two researchers, we reduced this threat by adopting SLR protocols \citep{ref:kitchenham2004procedures,ref:petersen2015guidelines}.

Publication bias is another threat, as the published research might only discuss positive aspects / projects, while negative or controversial aspects / results might not be published. Especially when considering advantages and drawbacks of testing processes, the positive reported research might heavily surpass negative research due to this bias---this however, might not reflect the real state in industry. Reading this SLR has to take into account the presence of this bias, as it is difficult to completely remove.

The quality of the sample of studies obtained with respect to the targeted population is another theoretical threat.  It is known that an SLR cannot cover the whole population of research published in the area, but rather provide a subset \citep{ref:kitchenham2004procedures}. We mitigated this threat by using different search queries for the search engines, as each one has its own peculiarities. This provided a better initial result-set to start with. We further decided to increase the quality of the selected studies by including only studies that met a certain threshold of reference citations.

\textbf{Descriptive Validity.} Descriptive validity deals with the accuracy of the facts reported by researchers based on the observed phenomenon \citep{ref:petersen2013worldviews}.

Poorly designed data extraction forms and recording of data can be a descriptive validity threat. In our case, we used the JabRef tool~(\url{http://www.jabref.org}) to store, manage, and annotate reference lists. Such tool was used for collaboration between researchers, tracking all the selection process. We believe this threat was in full control, as we could always go back to a previous step of the SLR process, if needed, and the whole reporting process was based on a single repositories for collaboration across researchers. About the extensiveness of the review, we did not consider the inclusion of grey literature~\cite{ref:garousi2016need} or the snowballing technique~\cite{ref:wohlin2014guidelines} to search related literature. The reason was mainly for time constraints, but also other challenges that are implied in grey literature, like the definition of a strategy for the identification of the credibility of sources~\cite{ref:benzies2006state}. 

\textbf{Interpretative (Construct) Validity}. Interpretative validity, deals with how much reliable are the conclusions that have been drawn from the evidence collected~\citep{ref:petersen2013worldviews}. 

Researchers bias can play a role also in this context, as interpretation of data might be steered according to previous beliefs and thoughts, with confirmation bias being one of such biases \citep{ref:nickerson1998confirmation-bias}. We believe this threat to validity is reduced in the current study, as the research protocol involved evaluation of the data and drawing of conclusions from the researchers involved.

\textbf{External Validity.} External validity deals with how much the findings could be generalized to other cases \citep{ref:petersen2013worldviews}.

In our context, the relevance would be on how much we can generalize empirical results from the reviewed articles to other industrial cases. In software engineering, there is high variability of results that depends on the context \citep{ref:petersen2009context,ref:dybaa2012works}. In this SLR, we reported the domain in which different testing processes were applied: we cannot ensure that results can be generalized even to other cases within the same domain: as such, readers need to take into account this threat when consulting the SLR findings.

\textbf{Repeatability / Reproducibility Validity.} Repeatability / reproducibility Validity deals with how much repeatable (reproducible) the performed process is. Reproducible research can be defined as any research that can be reproduced from the published materials by other researchers independently \citep{ref:madeyski-wider2017}. Throughout the study, we followed the concept of reproducible research \citep{ref:madeyski-reproducible2015, ref:madeyski-wider2017}. Following the SLR protocol \citep{ref:kitchenham2004procedures,ref:petersen2015guidelines} and reporting accurately all the steps was one way to increase the repeatability and reproducibility of the current study. Some parts, such as running the queries ex-post and all the parts that involve human judgement might involve challenges, but all the intermediary steps of research as stored in JabRef files will be available to other researchers. More advanced structured approaches to build search strings could be used, such as PICO (Patient/Intervention/Control/Outcome) \cite{ref:santos2007pico}. We believe this is a limited threat, as, in our case, we defined keywords from the research questions and used them for constructing search strings. To improve the results, we had several iterations trying different queries and optimizing them according to the search results.

\section{Conclusion}
Software testing is a key part of the software development process. The aim of this review was to provide an overview of existing software models that can be applied to improve the overall testing process. As the methodology for conducting the review, we used the Systematic Literature Review (SLR), to find empirical studies reporting benefits and drawbacks from the application of the models. Overall, 17 testing models were identified during the process. They differ in the area of usage and model representation. Most of the models are described as universally applicable, but several approaches were customized to satisfy specific requirements of domains such as embedded software, automated testing or military systems. We focused also on the aspects that the testing models helped to improve in the concrete cases. Most of the empirical studies describe the adaptation of a model to the testing process and thus standardization of the process as the main improvement. The next important supported aspect is the improvement of product quality and several studies focus on specific aspects such as measurements or defects detection / reduction. 

The adoption of a testing model can be influenced by advantages and drawbacks of the model under specific circumstances. Besides the specific domain or different improvement procedures, models provide several unique advantages. Based on the gathered information, we found that although most of the existing models are described as generally applicable, some organizations consider that the models are not suitable for small- to medium-sized enterprises and insufficient for specific domains such as defense or embedded software development. The models apply also different practices during the phases, which can be crucial for the model selection.

The review shows that there are many testing models with different characteristics that can fulfill the numerous requirements of organizations. However, it is important to select the appropriate approach based on the required objectives.  

\bibliographystyle{plainnat}

\end{document}